# NEW VIEWS OF CRYSTAL SYMMETRY

## ECKHARD HITZER

*Name*: Eckhard Hitzer, Physicist (b. Wolfratshausen, Germany, 1965).
*Address:* Department of Applied Physics, University of Fukui, 3-9-1 Bunkyo, Fukui, Japan.
*E-mail*: hitzer@mech.fukui-u.ac.jp; *Home-page*: http://sinai.mech.fukui-u.ac.jp/
*Fields of interest*: Clifford geometric algebra, Crystallography, Neural networks, Elementary particle physics, Signal Processing (Christian Faith and Science, Education, Travel).
*Publications*: Perwass & Hitzer, 2005; Hitzer & Perwass, 2009.

***Abstract:*** *Already Hermann Grassmann's father Justus (1829, 1830) published two works on the geometrical description of crystals, influenced by the earlier works of C.S. Weiss (1780-1856) on three main crystal forces governing crystal formation. In his 1840 essay on the derivation of crystal shapes from the general law of crystal formation Hermann established the notion of a three-dimensional vectorial system of forces with rational coefficients, that represent the interior crystal structure, regulate its formation, its shape and physical behavior. In the Ausdehnungslehre 1844 (§ 171) he finally writes: I shall conclude this presentation by one of the most beautiful applications which can be made of the science treated, i.e. the application to crystal figures (Scholz, 1996). The geometry of crystals thus certainly influenced the Ausdehnungslehre. In this paper we see how Grassmann's work influenced Clifford's creation of geometric algebras, which in turn leads to a new geometric description of crystal symmetry suitable for modern computer algebra graphics.*

Grassmann's work in turn influenced W.K. Clifford (1878) in England: *I propose to communicate in a brief form some applications of Grassmann's theory ... I may, perhaps, therefore be permitted to express my profound admiration of that extraordinary work, and my conviction that its principles will exercise a vast influence upon the future of mathematical science.* Conformal Clifford (geometric) algebra has in turn led at the beginning of the 20th century to a new fully geometric description of crystal symmetry in terms of so-called versors (Hestenes & Holt, 2007). Versors are simply (Clifford) geometric products of five-dimensional vectors conformally representing general planes in three-dimensional (3D) Euclidean space (by their 3D normal vector and the directed distance from the origin). Each plane's vector geometrically represents a reflection at the plane, the geometric products of several plane vectors represents the combination of reflections at the respective planes.

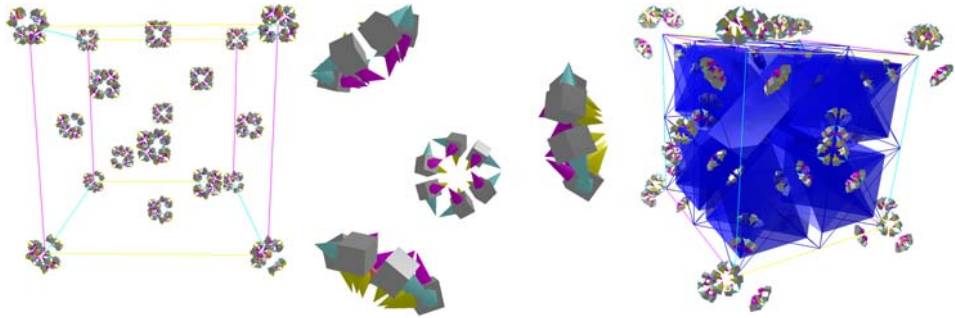

Figure 1: Left: Diamond cell in Space Group Visualizer. Center: 24 general elements in 3D showing diamond point symmetry of one vertex. Right: Diamond reflection planes.

As expected three crystal specific 3D vectors are enough to construct all symmetry versors of any type of crystal. With the geometric algebra capable graphics software CLUCalc this concept can be implemented in every detail, such that the abstract beauty of the enormously rich symmetry of crystals can be fully visualized by state-of-the-art 3D computer graphics: The Space Group Visualizer (SGV), a tailor-made CLUCalc Script (Perwass & Hitzer, 2005; Hitzer & Perwass, 2009). To be precise, the SGV is thus capable of showing every plane of reflection and glide-reflection symmetry, all axis of rotations, screw-rotations and rotoinversions, and every center of inversion. It further allows to dynamically visualize the action of any symmetry operation on a general element (representing atoms, molecules or ions). We thus have, 165 years after the Ausdehnungslehre of 1844, an explicit form of the beauty, which Grassmann may have had in mind, when he wrote eloquently: *one of the most beautiful applications.*

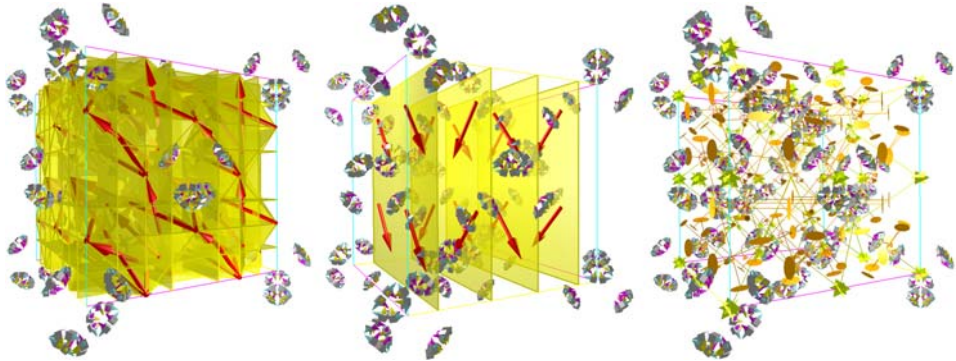

Figure 2: Left: All glide planes of a diamond cell. Centre: Pairs of diamond glide planes. Right: Symmetry rotation axis. Angles indicated by colours and arc segments.

*Geometrically* a diamond cell lattice (type: face centred cubic = fcc) is highly symmetric. That means there is an enormous variety of possible geometric transformations, that leave the lattice as a whole invariant, including all lengths and angles. These symmetry operations include single cell transformations that leave a cell vertex *point invariant*: planes of *reflections* (through the vertex), *rotations* (with axis trough the vertex, and *inversions* ($x \rightarrow -x$, centred at the vertex), and *rotoinversions* (inversions followed by a rotation). The 24 symmetry transformations of a diamond vertex point group create 24 symmetric copies of a general asymmetric element placed next to the invariant point, see Fig. 1 (left), or enlarged in Fig. 1 (center). In pure diamond one carbon atom is located at the centre of this cluster (plus one at 1/4 distance away along a cubic space diagonal).

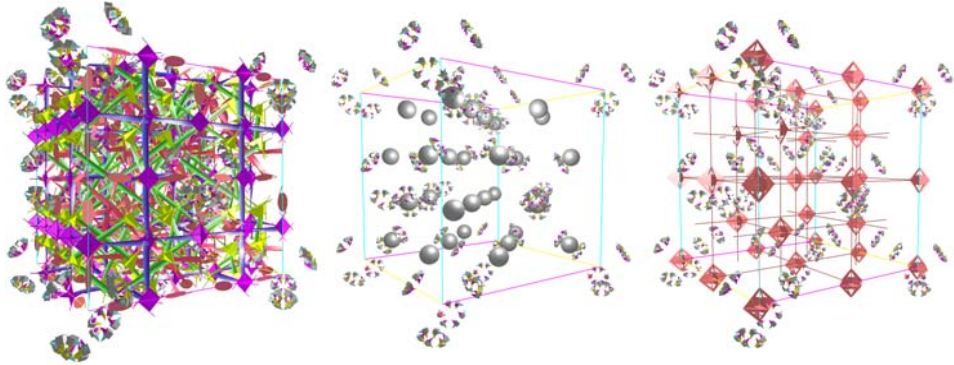

Figure 3: Left: All screw symmetry axis of diamond. Centre: All centres of inversion. Right: All rotoinversions passing through a single diamond lattice cell.

Figure 4: Left: All symmetries located in a single diamond cell. Right: Space group selection in the online Int. Tables of Crystallography (left half) and the SGV (right half).

The inclusion of integer lattice translations (from fcc vertex to fcc vertex) can lead to new planes of reflection, see Fig. 1 (right). The combination of a plane of reflection with a lattice translation not perpendicular to the plane leads to a combined *glide reflection*, see Fig. 2 (left), where (red) vectors indicate the parallel glide motion. The perpendicular translation component displaces the reflection plane in normal direction, and the parallel translation component creates a glide motion parallel to the plane. Pairs of characteristic diamond glides are shown in Fig. 2 (centre).

A sequence of two reflections at two planes results in a *rotation*, see Fig. 2 (right). This rotation has the intersection line of the two planes as its axis and twice the (dihedral) angle between the two planes is the resulting rotation angle. All the rotation axis seen in Fig. 2 (right) are lines of intersection of reflection planes of Fig. 1 (right). A lattice translation perpendicular to the rotation axis after a rotation, effectively creates another rotation also already contained in Fig. 2 (right). But if we perform a translation not normal to the rotation axis, with a translation component parallel to the rotation axis, we get a new transformation, a so-called *screw*. So a screw is a rotation followed by a translation along the screw axis, resulting in a directed helical motion around the screw axis, see Fig. 3 (left).

Combining an inversion with a subsequent lattice translation yields a new centre of inversion, see Fig. 3 (centre). The combination of an inversion with a rotation leads to a *rotoinversion*. Characteristic for the diamond lattice are the 90° rotoinversions depicted in Fig. 3 (right). The total graphical depiction of these symmetries in Fig. 4 (left) gives an idea of the intricate complexity of the symmetries possessed by the diamond lattice. The ITA (Hahn, 2005) depict the symmetries of diamond by showing a quarter of an orthographic 2D projection of a side of a cubic cell. The SGV allows to open an extra window with the ITA online and navigate synchronously in both, see Fig. 4 (right).

The author thanks his dear family, C. Perwass and M. Aroyo. He acknowledges God the creator: *I therefore believe the truths revealed in the Bible, not because they are written in the Bible, but because I have experienced in my own conscience their power of blessing, their eternal, divine truth* (Grassmann, 1878).